\documentclass[sn-nature]{sn-jnl} 

\usepackage{graphicx}%
\usepackage{multirow}%
\usepackage{amsmath,amssymb,amsfonts}%
\usepackage{amsthm}%
\usepackage{mathrsfs}%
\usepackage[title]{appendix}%
\usepackage{xcolor}%
\usepackage{textcomp}%
\usepackage{manyfoot}%
\usepackage{booktabs}%
\usepackage{algorithm}%
\usepackage{algorithmicx}%
\usepackage{algpseudocode}%
\usepackage{listings}%

\usepackage{xcolor} 

\definecolor{mygreen}{RGB}{0,100,0}

\theoremstyle{thmstyleone}%

\theoremstyle{thmstyletwo}%

\theoremstyle{thmstylethree}%

\makeatletter
\newcommand{\manualref}[2]{%
  \@namedef{r@#1}{{#2}{}{}{}{}}%
}
\makeatother 
\manualref{fig:SampleBalance}{Supplementary Fig. 1}
\manualref{fig:MCtest}{Supplementary Fig. 2}
\manualref{alterZ}{Supplementary Fig. 3}
\manualref{fig:TNG-map}{Supplementary Fig. 4}
\manualref{fig:sfr_azimuthal}{Supplementary Fig. 5}
\manualref{fig:alignment}{Supplementary Fig. 6}
\manualref{fig:sfr_scaling}{Supplementary Fig. 7}
\manualref{fig:TNG-pair}{Supplementary Fig. 8}
\manualref{fig:TNG-SampleBalance}{Supplementary Fig. 9}

\DeclareUnicodeCharacter{2500}{--}

\begin{document}

\title[Observational Evidence for Anisotropic Metal Excess around Galaxies]{Observational Evidence for Anisotropic Metal Excess around Galaxies}

\author[1,2]{\sur{Cheqiu Lyu}}

\author*[1,2]{\sur{Enci Wang}}\email{ecwang16@ustc.edu.cn}

\author[1,2]{\sur{Zeyu Chen}}
\author[1,2]{\sur{Chengyu Ma}}
\author[3,4]{\sur{Yangyao Chen}}
\author[1,2]{\sur{Haoran Yu}}
\author[1,2]{\sur{Xu Kong}}

\affil*[1]{\orgdiv{Department of Astronomy}, \orgname{University of Science and Technology of China}, \orgaddress{\city{Hefei}, \postcode{230026}, \state{Anhui Province}, \country{China}}}

\affil[2]{\orgdiv{School of Astronomy and Space Science}, \orgname{University of Science and Technology of China}, \orgaddress{\city{Hefei}, \postcode{230026}, \state{Anhui Province}, \country{China}}}

\affil[3]{\orgdiv{School of Astronomy and Space Science}, \orgname{Nanjing University}, \orgaddress{\city{Nanjing}, \postcode{210093}, \state{Jiangsu Province}, \country{China}}}

\affil[4]{\orgdiv{Key Laboratory of Modern Astronomy and Astrophysics}, \orgname{Nanjing University, Ministry of Education}, \orgaddress{\city{Nanjing}, \postcode{210093}, \state{Jiangsu Province}, \country{China}}}

\abstract{The exchange of matter and energy between galaxies and their surroundings drives the cosmic baryon cycle, yet mapping metal transport remains an observational challenge. While simulations predict that galactic winds escape anisotropically along minor axes, evidence for chemical enrichment in neighboring galaxies is limited. We analyze 1,433 galaxy pairs from the Dark Energy Spectroscopic Instrument survey and detect a gas-phase metallicity excess of 14.6\% $\pm$ 3.7\% to 24.2\% $\pm$ 2.6\% in neighbors aligned with the minor axis of massive primary at projected separations of 15--60 kpc, qualitatively consistent with IllustrisTNG simulation. Permutation tests indicate that this excess varies from a marginal detection ($>$92\% confidence) at closer separations (15--30 kpc) to a significant signal ($>$98\% confidence) at intermediate separations (30--60 kpc). Here, we show that these results provide empirical constraints on galactic feedback models and are consistent with a scenario where anisotropic outflows can enrich nearby galaxies, a mechanism that complements other environmental processes.}


\maketitle

\section*{Introduction}

Galactic outflows, driven by stellar feedback and active galactic nuclei (AGN), are required to regulate the star formation in galaxies, and reproduce the fundamental stellar mass–halo mass ($M_*-M_h$) relation in both semi-analytic models and cosmological hydrodynamical simulations \cite{White1978, White1991, Veilleux2005, Murray2005, Behroozi2013, Moster2013, Hopkins2014, Somerville2015, Heckman2017, Lyu2023, Mo2024, Li2025, Zhao2025,WangK2025, WangK2026a}. Beyond this primary regulatory role, outflows are also the main mechanism for enriching the circumgalactic and intergalactic medium (CGM and IGM) with heavy elements \cite{Oppenheimer2006, Tumlinson2017, Peroux2020, Veilleux2020, Wang2022a, Wang2022b, Faucher-Giguere2023, Ma2024, ChenYY2025, Yu2026, Lyu2026, WangK2026b}. A critical aspect of these outflows is their inherent anisotropy. For decades, observations of local starbursts and cosmological simulations have consistently indicated that feedback energy and enriched material escape preferentially along the path of least resistance---perpendicular to the galactic disk---forming a bipolar or conical structure \cite{Heckman1990, Muratov2015, Nelson2019, Guo2023}. 

Observational efforts, using a variety of complementary techniques, have provided compelling evidence for this anisotropic picture. Stacking analyses of background quasar spectra show that the column density of cool gas, traced by Mg~II absorption, is significantly enhanced along the minor axes of star-forming galaxies \cite{Chen2025}. In parallel, direct imaging of the CGM in emission has revealed similar anisotropic structures, both in the form of extended bipolar cones of Mg~II emission around high-redshift galaxies \cite{Guo2023} and in the angular dependence of optical emission-line ratios (e.g., [N~II]/H$\alpha$), which trace the gas ionization state \cite{Zhang2024}. One pioneering absorption-line study even used dust depletion ([Zn/Fe]) as a proxy to argue that the gas along the minor axis is more metal-rich \cite{Wendt2021}. Taken together, these studies build a strong consensus that outflows create an anisotropic CGM, with a greater column density of cool, metal-loading gas and a distinct ionization state along the minor axis. Gas-phase chemical evolution of galaxies has been extensively mapped since the early era of the Sloan Digital Sky Survey (SDSS) \cite{Tremonti2004, Ellison2008, Ellison2009,Mannucci2010}. These studies established that satellite galaxies often exhibit distinct metallicity patterns compared to centrals due to various environmental processes \cite{Cooper2008, Pasquali2010,Scudder2012,Peng2014,WangK2023}. However, direct evidence for an anisotropic metallicity excess around galaxies---specifically one manifested in the interstellar medium (ISM) of neighboring galaxies---is still lacking.

\begin{figure}[ht]
\centering
\includegraphics[width=1\textwidth]{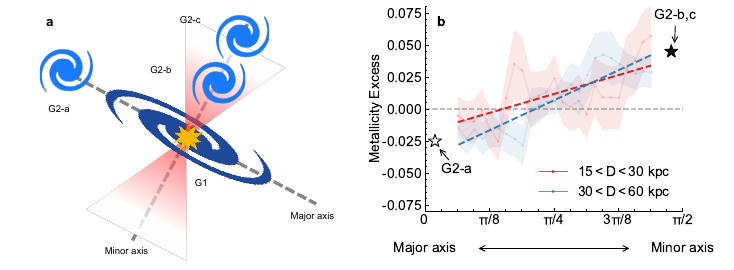}
\caption{\textbf{Schematic illustration and observational evidence of the anisotropic metal excess hypothesis.} \textbf{a} A star-forming primary galaxy (G1) drives a bipolar, metal-enriched outflow (red cones) along its minor axis, the path of least resistance. Neighboring galaxies (G2) located along G1's major axis (e.g., G2-a) are largely unaffected by this outflow. In contrast, neighbors residing along the minor axis (e.g., G2-b, G2-c) are expected to be more significantly influenced by the metal-rich gas. \textbf{b} Empirical examination of this hypothesis using our DESI galaxy pair sample ($15 < D < 60$~kpc). The running median of the metallicity excess ($\Delta\log(\text{O/H})_{\text{G2}}$) shows a clear positive correlation with the azimuthal angle ($\theta$), consistent with the scenario where neighbors near the minor axis are systematically enriched. Open and solid stars represent the characteristic regimes of metallicity depletion and excess, respectively, mapping the idealized configurations of neighboring galaxies (G2-a, G2-b, and G2-c) from panel \textbf{a} onto the observed statistical trends. Shaded regions represent $1\sigma$ bootstrap uncertainties, and the dashed lines denote linear fits to the running medians. This figure shows that the observed azimuthal dependency is consistent with the geometric influence of galactic outflows. Source data are provided as a Source Data file.}\label{carton}
\end{figure}

This anisotropic geometry leads to a clear and testable prediction regarding the chemical environment around galaxies, as illustrated in Fig.~\ref{carton}. A primary, star-forming galaxy (G1) expels metal-enriched gas predominantly along its minor axis. Neighboring galaxies (G2) positioned within this outflow cone (such as G2-b or G2-c) are expected to be exposed to, and may potentially accrete, this metal-rich material. This process could elevate the neighbor's own gas-phase metallicity relative to their counterparts of similar stellar mass. This chemical enrichment translates into a distinct observational signature on the mass-metallicity relation (MZR). While an unenriched neighbor located along the primary's major axis (G2-a) would follow the standard MZR, a neighbor enriched by the outflow is expected to exhibit a positive metallicity offset at a given stellar mass ($\Delta \log(\text{O/H}) > 0$). 

In this work, we present a test of this scenario by leveraging the gas-phase metallicity of neighboring galaxies as a direct chemical probe. Using a large sample of galaxy pairs, we present the statistical evidence that metallicity is enhanced in neighboring galaxies aligned with the minor axes of their massive companions, supporting a scenario in which anisotropic outflows enrich nearby galaxies and drive anisotropic metal transport. This provides an observational constraint on the metal-loading and propagation of galactic feedback.

\section*{Results}

\subsection*{Sample Selection and Geometry}

\begin{figure}[ht]
\centering
\includegraphics[width=1\textwidth]{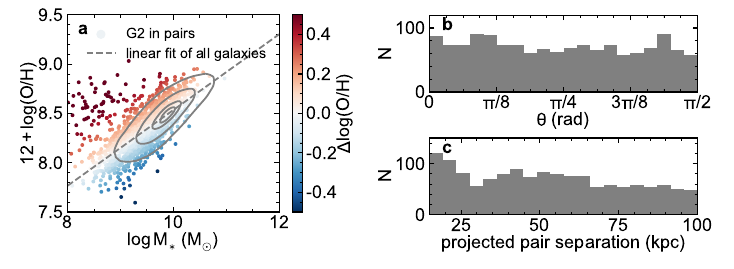}
\caption{\textbf{Sample selection and properties of galaxy pairs.} \textbf{a} The mass-metallicity relation for the parent sample (contours) and selected neighboring galaxies (G2) in pairs (points). The color of each point indicates its metallicity offset, $\Delta\log(\text{O/H})$, from the fitted linear relation of the parent sample, which is shown as a dashed grey line. Grey contours represent the density distribution of the parent sample, with levels enclosing 16\%, 50\%, 84\%, and 95\% of the total distribution. \textbf{b} The distribution of the azimuthal angle ($\theta$) between the primary galaxy (G1)'s major axis and the vector to its neighbor (G2). \textbf{c} The distribution of the projected pair separation in kiloparsecs (kpc). Source data are provided as a Source Data file.}\label{MZR}
\end{figure}

We construct a sample of galaxy pairs using Dark Energy Spectroscopic Instrument (DESI)\cite{Levi2013,DESICollaboration2022,Levi2019,Silber2023,Miller2024,DESICollaboration2024b} catalog (see \hyperref[sec:method_pair]{Methods, subsection Galaxy Pair Identification and Final Sample}). Fig.~\ref{MZR} provides an overview of the fundamental properties and geometric configuration of this sample. Fig.~\ref{MZR}\textbf{a} presents the MZR for our sample galaxies. This relation demonstrates a strong positive correlation where more massive galaxies are, on average, more chemically evolved and thus more metal-rich. We determine the gas-phase metallicity ($12+\log(\text{O/H})$) for each galaxy using the N2S2H$\alpha$ calibrator Dopita et al. (2016)\citep{Dopita2016} as our fiducial diagnostic, as it is largely insensitive to reddening and the ionization parameter \citep[e.g.,][]{Zhang2017, Hwang2019} (see \hyperref[sec:method_metallicity]{Methods, subsection Gas-phase Metallicity Measurement}). A key parameter for our study is the metallicity offset, $\Delta\log(\text{O/H})$, which quantifies the metallicity deviation of a galaxy from the best-fit linear relation of the parent sample. Galaxies that are metal-rich for their mass exhibit positive $\Delta\log(\text{O/H})$ (redder colors), while those that are metal-poor show negative values (bluer colors).

The geometric properties of the pairs are characterized in Fig.~\ref{MZR}\textbf{b,c}. They exhibit a uniform distribution in the azimuthal angle $\theta$, measured between the major axis of G1 and its neighbor (from 0 to $\pi/2$ radians, see \hyperref[sec:method_geometry]{Methods, subsection Galaxy Orientation and Geometry}). The projected pair separation covers the full range from 15 to 100~kpc, with a moderate decrease in the number of pairs at larger separations. This lack of any significant selection bias in orientation ensures that our analysis is not dominated by specific pair configurations and that any detected anisotropy in metallicity is a real physical effect rather than an observational artifact (see \ref{fig:SampleBalance} and \hyperref[sec:method_balance]{Methods, subsection Sample Balance and Selection Effects}).

\subsection*{Observational Evidence}

\begin{figure}[!ht]
\centering
\includegraphics[width=\textwidth]{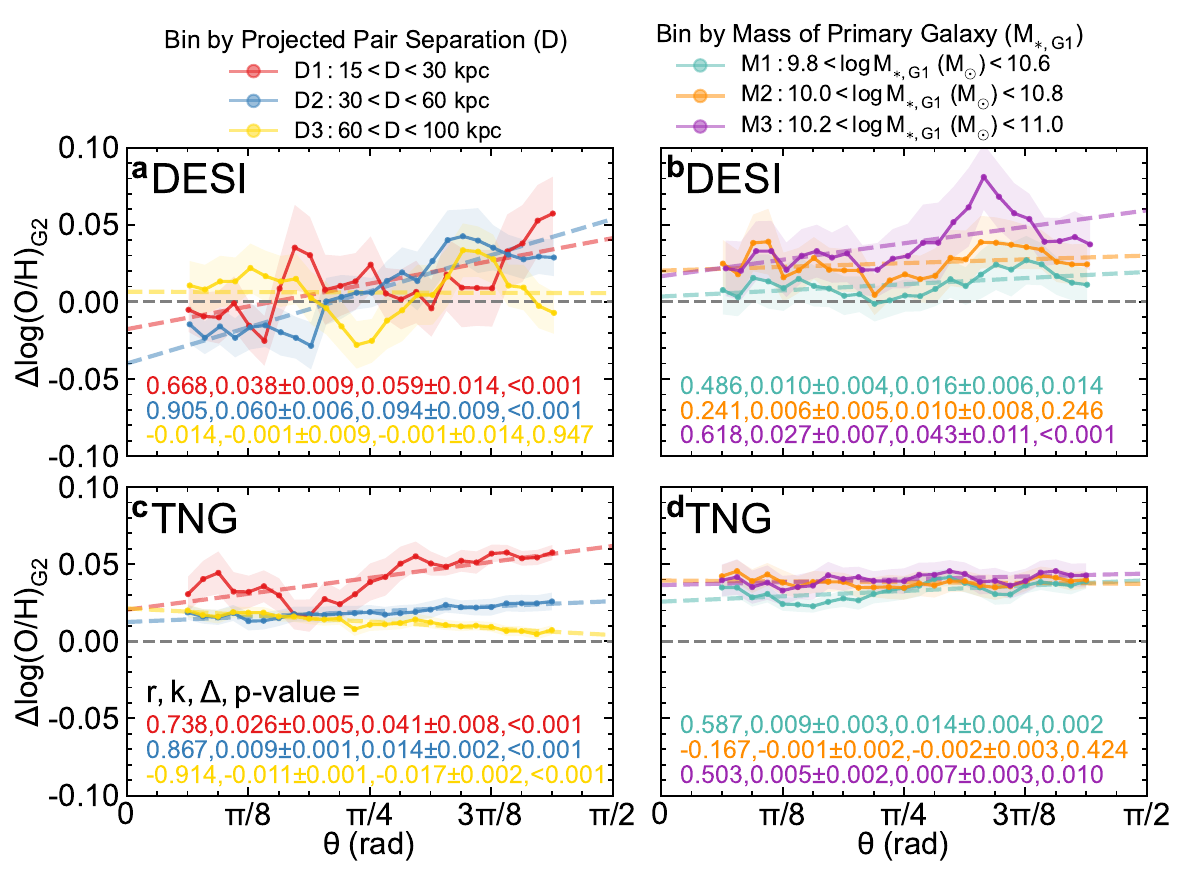}
\caption{\textbf{Anisotropic metal excess in galaxy pairs from observations and simulations.} The running median of the metallicity offset of the secondary galaxy (G2, $\Delta\log(\text{O/H})_{\text{G2}}$) as a function of the azimuthal angle ($\theta$), relative to the major axis of the primary galaxy (G1). A value of $\theta=0$ corresponds to the neighbor residing along the major axis, while $\theta=\pi/2$ aligns with the minor axis. \textbf{a,b} show observational results from our DESI sample, while \textbf{c,d} show corresponding results from the TNG cosmological simulation. \textbf{a,c} The sample is divided into three bins of projected pair separation ($D$). A clear positive trend is observed for the closest pairs (D1: $15 < D < 30$~kpc, red; and D2: $30 < D < 60$~kpc, blue), indicating higher metallicity for neighbors along the primary's minor axis. For widely separated pairs (D3: $60 < D < 100$~kpc, gold), the minor-axis enhancement is no longer evident, serving as a physical control sample. \textbf{b, d} The sample is binned by the stellar mass of the primary galaxy ($M_{\text{*,G1}}$). The anisotropic signal is most pronounced for pairs with the most massive primaries (M3: purple). The shaded regions represent the $1\sigma$ standard error of the running medians, calculated via bootstrap resampling. The thick dashed lines represent linear fits to the running median points. Statistical metrics for each fit are annotated in the panels, including the Pearson correlation coefficient ($r$), the slope ($k$), the extrapolated total metallicity enhancement from the major to the minor axis ($\Delta$), and the $p$-value. The qualitative agreement between the observed trends in DESI and the predictions from the TNG simulation, characterized by the consistent presence of a positive angular gradient ($k>0$) for closer pairs and its shared dependence on primary galaxy mass, provides plausible evidence that metal-enriched galactic outflows are likely to enrich neighboring galaxies. While the simulation captures these key physical trends, the simulated angular slopes are shallower than those observed (see Discussion). Source data are provided as a Source Data file.}\label{theta-dZ}
\end{figure}

The main result is presented in Fig.~\ref{theta-dZ}, which provides evidence for anisotropic metal excess between galaxies. We plot the median metallicity offset of the secondary galaxy ($\Delta\log(\text{O/H})_{\text{G2}}$) as a function of its azimuthal angle with respect to the major axis of the primary galaxy (G1). This framework allows us to test the hypothesis that metal-enriched outflows, typically launched perpendicular to the galactic disk, can chemically influence a neighboring galaxy. A positive correlation in this plane signifies higher metallicity excess along the primary's minor axis ($\theta = \pi/2$) compared to its major axis ($\theta = 0$).

In the DESI sample (Fig.~\ref{theta-dZ}\textbf{a,b}), using a running median approach with bootstrap error estimation (see \hyperref[sec:method_statistics]{Methods, subsection Statistical Analysis of Azimuthal Dependence}), we find a positive correlation between $\Delta\log(\text{O/H})_{\text{G2}}$ and azimuthal angle ($\theta$) for closer pairs ($15 < D < 60$~kpc, $p < 0.001$), Specifically, the Pearson correlation coefficient ($r$) reaches 0.668 for the $15<D<30$ kpc bin and 0.905 for the $30<D<60$ kpc bin, suggesting a dependence of the neighbor's metallicity enhancement on its angular position relative to the primary galaxy's disk. To assess the statistical significance of this correlation and evaluate its sensitivity to random data fluctuations or the binning procedure, we perform Monte Carlo permutation tests on both the running medians and the unbinned scatter data points
(see \ref{fig:MCtest} and \hyperref[sec:method_statistics]{Methods, subsection Statistical Analysis of Azimuthal Dependence}). 
We find that the statistical significance of the anisotropic signal depends on the pair separation. For the closest pairs ($15 < D < 30$ kpc), we report a marginal detection of anisotropic excess (confidence level (CL) $> 92\%$), translating to a total enhancement of $\Delta = 0.059 \pm 0.014$ dex (14.6\% $\pm$ 3.7\%, calculated as 
$10^{\Delta}-1$ with errors propagated accordingly) from the major axis to the minor axis. Here, the CL is defined as the probability that the observed angular gradient is not a random occurrence, calculated as the fraction of randomized realizations yielding a slope less extreme than the observed 
slope in the direction of the detected trend. In contrast, this anisotropic signal becomes significant for pairs at intermediate separations ($30 < D < 60$ kpc), where the permutation test yields a CL $>98\%$ and the enhancement reaches $\Delta = 0.094 \pm 0.009$ dex (24.2\% $\pm$ 2.6\%). For widely separated pairs ($60 < D < 100$ kpc), the trend is completely flat, effectively serving as a physical control sample, and suggesting the spatial limitation of the metal-enriched outflows.

Furthermore, Fig.~\ref{theta-dZ}\textbf{b} reveals that the metallicity excess pattern depends on the primary galaxy's mass. With the quantitative fits, we clarify that the influence of the most massive primaries (M3: $10.2 < \log(M_{\text{*,G1}}/\mathrm{M_\odot}) < 11.0$, purple) manifests in two distinct ways. First, they produce a higher global metallicity excess across all angles, as noted by the overall upward shift of the trend. Second, they exhibit the steepest positive angular dependence (slope $k = 0.027 \pm 0.007$, $p = 0.001$). Permutation tests on the scatter data points support the presence of this steeper gradient, reaching a CL of approximately $95.4\%$ (see \ref{fig:MCtest} and \hyperref[sec:method_statistics]{Methods, subsection Statistical Analysis of Azimuthal Dependence}). In contrast, for less massive primaries (e.g., cyan), the angular gradient is notably shallower ($k = 0.010 \pm 0.004$). This is consistent with the scenario in which more massive galaxies, which possess deeper potential wells and higher intrinsic gas-phase metallicities, drive more heavily enriched outflows that imprint a more directional chemical signature on their environments. Importantly, these trends are not an artifact of our fiducial metallicity diagnostic; we have tested them using four additional calibrators and find a qualitatively consistent signal in most of the cases (see \hyperref[sec:method_metallicity]{Methods, subsection Gas-phase Metallicity Measurement} and \ref{alterZ}).

\subsection*{Comparison with IllustrisTNG}

We compare our results with predictions from the Illustris The Next Generation \citep[TNG,][]{Nelson2018,Pillepich2018} cosmological hydrodynamical simulation (Fig.~\ref{theta-dZ}\textbf{c,d} and see \hyperref[sec:method_tng]{Methods, subsection Galaxy Pairs in the IllustrisTNG Simulation}). Qualitatively, the TNG simulation reproduces the key trends seen in the DESI data: (1) a statistically positive correlation ($k>0, p<0.001$) for pairs with $15<D<60$~kpc; (2) the disappearance of this signal ($k$ about 0) at larger separations ($D>60$~kpc); and (3) the intensification of the metallicity excess in pairs with more massive primaries. Quantitatively, the simulated signal appears shallower than that observed. To test the robustness of this subtle trend, we perform the same Monte Carlo permutation tests on the TNG sample as we do for the DESI data. Despite the smaller slope for close pairs, the positive correlation yields CL ranging from marginal to significant (CL approximately 91\%--99\%, depending on the separation bin D1 or D2 and fitting method; see  \hyperref[sec:method_statistics]{Methods, subsection Statistical Analysis of Azimuthal Dependence} and \ref{fig:MCtest}). This demonstrates that the qualitative agreement is statistically grounded under permutation testing.

However, we caution against a direct quantitative comparison of these amplitudes. As shown in \ref{alterZ}, the magnitude of the observed metallicity variation depends on the choice of calibration method. Similarly, simulated metallicities are subject to uncertainties in nucleosynthetic yields and sub-grid feedback models. Notably, the discrepancy in CL between the TNG and DESI samples---where the simulation yields a more marginal detection (CL approximately 91\%--94\% in Bin D2) compared to the highly significant observational signal (CL approximately 98\%--99.9\%)---further underscores this difference in amplitude, as shallower slopes inherently result in lower CLs during permutation tests. The simulated neighbors exhibit a globally positive metallicity offset ($\Delta \log(\text{O/H}) > 0$) nearly across all angles. This reflects the efficient metal enrichment of the CGM and intragroup medium in the TNG model, which produces high mass-loading, metal-enriched outflows that recycle heavy elements into the local halo environment~\cite{Nelson2019, Peroux2020b, Weng2024, Ramesh2023}. Consequently, neighboring galaxies in these enriched halos may accrete gas with a higher baseline metallicity than typical field galaxies. The anisotropic trend we observe is thus a directional enhancement superimposed on this globally enriched background.

The gas environment surrounding these galaxies suggests a differential dilution process. While all neighbors undergo metallicity dilution through the accretion of pristine cosmic web gas, those along the minor axis are exposed to a persistent standing reservoir of metal-enriched gas ($\rm 12+\log(O/H) \gtrsim 8.5$ launched at velocities of approximately 40--80 $\mathrm{km\ s^{-1}}$; see \hyperref[sec:method_map]{Methods, subsection Velocity and Metallicity Map around Galaxies in IllustrisTNG} and \ref{fig:TNG-map}). Since the CGM gas is generally more metal-poor than the ISM of the neighbors, its accretion typically leads to a dilution of the neighbor's metallicity. However, as this reservoir along the minor axis may provide an enriched supply that mitigates the dilution typically caused by pristine inflows, the resulting dilution can be mitigated compared to major-axis counterparts. Consequently, the observed metallicity excess may arise from the net balance between this anisotropic enrichment and global dilution.

Given these systematic differences in absolute abundance scales, the simulation offers a plausible physical scenario that is consistent with the positive correlation. However, the amplitude of the angular trend appears modestly stronger in the DESI observations than in the simulation. This subtle discrepancy might indicate that the stellar feedback implemented in TNG is slightly less efficient at anisotropically enriching the local CGM compared to real galaxies. This overall agreement is not coincidental; A direct visualization of the gas environment around disk galaxies (see \hyperref[sec:method_map]{Methods, subsection Velocity and Metallicity Map around Galaxies in IllustrisTNG} and \ref{fig:TNG-map}) shows that the gas reservoir where neighbors locate is systematically more metal-rich along the minor axis compared to the major axis. This provides a plausible physical foundation for the differential dilution scenario by showing that the available gas supply is significantly more enriched along the outflow axis. This powerful, bipolar outflow that transports metal-enriched gas preferentially along the galaxy's minor axis, creating a stable chemical gradient that is likely to imprint itself on the ensemble population of neighboring galaxies before their orbital motion can randomize the signal. Our detection of an anisotropic metallicity excess is consistent with TNG predictions that metal-rich winds are channeled anisotropically~\cite{Weng2024}, providing a complementary approach that mitigates some of the uncertainties inherent in instantaneous absorption-line measurements~\cite{Weng2023}.

\section*{Discussion}
\subsection*{Nature and Dilution of the Metal Excess Signal}
While our analysis reveals a systemic trend of anisotropic metallicity excess, the modest amplitude of the signal representing a secondary but systematic environmental modulation on top of the dominant mass-driven metallicity, suggests that the observed effect may be diluted by a combination of observational and physical factors. Firstly, projection effects are inherent in any pair sample selected from a 2D sky survey. Although DESI provides precise spectroscopic redshifts with velocity uncertainties of approximately 30 km s$^{-1}$, the projected pair separation $D$ is only a lower limit on the true three-dimensional separation. Due to peculiar velocities, galaxies with small redshift differences may still be separated by large distances along the line of sight, which would naturally weaken the observed correlations \cite{Patton2013, Scudder2012}. Secondly, galactic outflows are not continuous but are episodic, with a finite duty cycle tied to bursts of star formation or AGN activity \cite{Hopkins2014, Muratov2015}. We are therefore observing a snapshot in time where only a subset of primary galaxies are actively driving a powerful, metal-enriched wind toward their neighbors. Thirdly, the process of enrichment is not instantaneous; there is a time delay between the launch of the outflow, its travel to the neighbors, and the subsequent mixing of the accreted material into the neighbor's star-forming interstellar medium. Finally, our fixed-aperture fiber measurements primarily probe the central regions of neighboring galaxies, where the metallicity is dominated by the galaxy's own in-situ star formation. Environmental enrichment is thus a secondary effect and is expected to be far more pronounced in the gaseous outskirts \cite{Sancisi2008, Tumlinson2017}. Therefore, the signal we detect is likely a muted signature of a enrichment event occurring in the CGM. While strict selection criteria for metallicity and orientation diagnostics limit the total sample size, the resulting dataset provides the high precision necessary to resolve these subtle environmental trends.

Despite potential diluting effects, the detection of a coherent anisotropic signal points toward a persistent enrichment mechanism. The observed trends are consistent with a standing reservoir of metal-rich gas maintained along the primary galaxy's minor axis (\ref{fig:TNG-map}), rather than an enrichment process that operates faster than the satellite's orbital crossing time. While orbital motion eventually moves satellites out of the minor-axis cone---a process that tends to mix the population and dilute the angular dependence---the presence of a statistically significant signal suggests that the chemical contrast between the axes is sufficiently strong to imprint a chemical signature on the ensemble population, despite the blurring effects of orbital dynamics.

\subsection*{Gas Stripping}

Gas stripping presents an alternative mechanism for metallicity enhancement. As demonstrated in TNG simulations, satellites interacting with dense galactic winds can undergo ram pressure stripping \cite{Nelson2019}. The relative importance of gas stripping versus accretion in these environments remains a subject of active debate. Since galaxies typically exhibit negative metallicity gradients, the preferential removal of loosely bound, low-metallicity gas from the outskirts could increase the integrated metallicity of the remaining system. However, our DESI fiber measurements primarily sample the inner regions (median coverage approximately 1.0 effective radius), the stripping of gas at large radii would likely have a secondary effect on our measured metallicities compared to the direct interaction with the primary's enriched gas. Furthermore, although stripping depends on the neighbor’s orbit and the local environment, it is expected to be largely isotropic relative to the primary disk's orientation. Therefore, stripping alone is unlikely to be the primary driver of the observed azimuthal dependence.

\subsection*{Interaction-Induced in-situ Star Formation}

We further evaluated whether the observed anisotropic metallicity excess could be driven by internal processes triggered by galaxy interactions rather than external accretion. Tidal interactions can drive radial gas inflows that typically dilute central metallicity \citep[e.g.,][]{Ellison2008, Rupke2010, Torrey2012}, though intense interaction-induced starbursts may eventually lead to re-enrichment \citep{Ellison2013}. To test this possibility, we analyze the azimuthal distribution of the star formation rate (SFR) for the secondary galaxies (see \hyperref[sec:method_sfr]{Methods, subsection Azimuthal Distribution of SFR Offset} and \ref{fig:sfr_azimuthal}). While we find a global SFR enhancement (by approximately 0.08--0.15 dex) for the closest pairs ($15 < D < 30$~kpc), this boost appears to be azimuthally decoupled from the metallicity signal. Quantitatively, permutation tests indicate a tentative negative angular gradient (e.g., $k_{\text{SFR}} = -0.071 \pm 0.020, p=0.002$ for Bin D1; \ref{fig:sfr_azimuthal}\textbf{b}), and the statistical significance is low to marginal (CL approximately 86\%--91\%).

Given the limited significance levels, they point toward a distinct spatial anti-correlation: the SFR enhancement peaks along the major axis and decreases toward the minor axis, whereas the metallicity excess is strongest along the minor axis. This decoupling disfavors in-situ chemical evolution as the primary mechanism, although it may still contribute to the global metallicity enhancement. Dynamically, there is no expectation for interaction-induced star formation to be maximized specifically along the primary’s minor axis. While an indirect channel---whereby outflow-induced pressure or turbulence triggers star formation---is physically plausible, it operates on timescales of about 300--500 Myr ~\cite{Zubovas2013,Bieri2016}, significantly longer than the direct transport of metal-rich gas ($< 100$ Myr~\cite{Muratov2015,Heckman2017b}). The lack of enhanced star formation within the minor-axis outflow cones (\ref{fig:sfr_azimuthal}) is consistent with the interpretation that such indirect mechanism may not be a primary contributor to the observed signal.

\subsection*{Shared Large-scale Structure and Cosmic Filaments}
Large-scale structure (LSS) or cosmic filaments may also influence the observed azimuthal signal. Pristine gas accretion from the cosmic web typically aligns with a galaxy’s major axis \cite{Danovich2015, Stewart2017}, which is consistent with the slightly negative metallicity offsets ($\Delta \log(\text{O/H}) < 0$) observed near $\theta = 0$. Therefore, LSS accretion does not contradict our findings; rather, it provides a physical mechanism for the low-metallicity baseline along the major axis.

To evaluate whether the minor-axis metallicity excess is consistent with being primarily driven by feedback, we perform two additional tests. First, orientation analysis shows no preferred spin alignment between pair members (Kolmogorov–Smirnov (KS) text $p = 0.8302$, \ref{fig:alignment}), suggesting the signal is not a geometric bias induced by coherent formation within a single filament. Second, for the closest pairs ($15 < D < 30$ kpc), the neighbor's metallicity excess shows a suggestive correlation with the primary’s SFR (\ref{fig:sfr_scaling}). While modest, this trend is qualitatively consistent with a feedback-driven scenario. The weak scaling likely reflects the inherent time-lag between gas expulsion and its eventual accretion; the instantaneous SFR may not fully capture the cumulative impact of past star-formation episodes (e.g., about 100~Myr ago) that drive the observed metallicity enhancement.

While these diagnostic tests do not fully preclude secondary contributions from other environmental processes, they collectively suggest anisotropic galactic outflows as the primary driver of the observed metal transport. The azimuthal metallicity pattern appears to be shaped by the interplay of two mechanisms: LSS-driven dilution along the major axis and feedback-driven enrichment along the minor axis.

In summary, the qualitative agreement between DESI observations and IllustrisTNG simulations supports a scenario in which galaxies chemically enrich their neighbors through a cross-pollination process likely mediated by anisotropic outflows. By isolating the directional transport of metals from the overall distribution of circumgalactic gas, these findings provide empirical constraints on the metal-loading and propagation of galactic feedback. While our diagnostic tests suggest that alternative mechanisms---such as interaction-induced star formation or large-scale structure accretion---are unlikely to be the primary driver of the observed azimuthal signal, they do not fully preclude secondary contributions from these processes. Definitive demonstration of the kinematic assimilation of this circumgalactic metal excess will require future simulation-based studies utilizing gas tracer particles. Ultimately, this work offers a data-driven basis to refine sub-grid feedback prescriptions in next-generation galaxy formation models.

\backmatter

\section*{Methods}
\subsection*{Cosmology}
To ensure a strictly fair comparison with the IllustrisTNG simulation, we adopt the Planck 2015 cosmology~\cite{PlanckCollaboration2016} for all distance-dependent calculations in our observational sample. The parameters are $\Omega_m = 0.3089$, $\Omega_\Lambda = 0.6911$, and a dimensionless Hubble parameter $h = 0.6774$.

\subsection*{DESI Y1 Catalog}

Our analysis is based on the first-year data release (Y1) of the DESI, a Stage-IV survey designed to probe cosmic acceleration by mapping the large-scale structure of the Universe \cite{Levi2013, DESICollaboration2022}. Utilizing a focal plane with 5,000 robotic fiber positioners, DESI provides the capacity for high-density, large-scale spectroscopic observation \cite{Levi2019, DESICollaboration2022, Silber2023, Miller2024}. The survey primarily targets four classes of extragalactic sources: quasars (QSOs), and three distinct galaxy populations known as the Bright Galaxy Survey (BGS), Luminous Red Galaxies (LRGs), and Emission Line Galaxies (ELGs).

The Y1 data release comprises spectra for over 18 million unique targets observed between May 2021 and June 2022, creating the most extensive three-dimensional map of the Universe to date over a footprint of 9,000 square degrees \cite{DESICollaboration2025}. The immense statistical power of this dataset has already yielded a generation of precision measurements constraining the cosmic expansion history from $z=0$ to 2 \cite{Adame2025, DESICollaboration2024b, Adame2025a, DESICollaboration2024c, Adame2025b}.

\subsection*{Parent Sample Construction}
\label{sec:method_sample}
We select a large, representative parent sample of star-forming galaxies from the BGS of the DESI Y1 catalog, imposing a redshift range of $0.05 < z < 0.49$. This redshift range ensures that the average 1.5-arcsecond diameter fiber aperture~\cite{DESICollaboration2022, DESICollaboration2024c} subtends a significant physical area, corresponding to a projected diameter of approximately 1.51~kpc at $z=0.05$ to approximately 9.33~kpc at $z=0.49$. We calculate the coverage of the fiber relative to the effective radius ($R_{\rm e}$, derived from the \texttt{SHAPE\_R} parameter in the imaging catalog) of each galaxy and find a median coverage of $R_{\rm fiber}/R_{\rm e}$ about 1 (see \ref{fig:SampleBalance}\textbf{f}). This implies that our spectra typically capture the integrated flux from the inner about $50\%$ of the star-forming disk. Moreover, the H$\alpha$ emission line falls well within the DESI spectral coverage. The sample is further restricted to galaxies with stellar masses greater than $10^8\ \mathrm{M_\odot}$. Since DESI BGS is a magnitude-limited survey ($r \lesssim 19.5$)~\cite{Hahn2023}, the sample is not volume-complete down to this mass limit across the entire redshift range. However, as this selection bias is isotropic with respect to galaxy orientation, it does not affect our analysis of angular dependence. As shown in \ref{fig:SampleBalance}, for selected pairs, the major-axis and minor-axis subsamples are well-balanced in stellar mass, redshift, and other observational parameters, indicating that our detected anisotropic metallicity excess is a physical effect rather than a selection artifact.

To ensure reliable metallicity diagnostics, we require a signal-to-noise ratio (S/N) $> 3$ in the five principal strong emission lines: H$\alpha\lambda 6563$, H$\beta\lambda 4861$, [N~II]$\lambda 6583$, [O~III]$\lambda 5007$, and the [S~II]$\lambda\lambda 6716, 6731$ doublet. Emission line fluxes for each galaxy are obtained from the DESI catalog. The fluxes are first corrected for interstellar reddening using the Balmer decrement method, assuming Case B recombination with an intrinsic H$\alpha$/H$\beta$ ratio of 2.86 \citep{Osterbrock2006} and a Cardelli et al. (1989)\citep{Cardelli1989} extinction law with $R_V = 3.1$. Galaxies with emission line ratios dominated by AGN are identified and removed using the BPT diagnostic diagram \cite{Baldwin1981, Kewley2001, Kauffmann2003}. This process resulted in a final parent sample of 376,083 star-forming galaxies.

\subsection*{Galaxy Pair Identification and Final Sample}\label{sec:method_pair}
From the parent sample, we construct a catalog of galaxy pairs based on their proximity in both projected pair separation and line-of-sight velocity. A pair is identified if both galaxies are present in our parent sample and satisfies a projected pair separation of $15 < D < 100$~kpc and a line-of-sight velocity difference of $|\Delta v| < 500\ \text{km s}^{-1}$, where $\Delta v = c\Delta z / (1+z)$. The lower pair separation threshold of 15~kpc is chosen to specifically exclude systems undergoing direct tidal disruption or mergers.

Within each identified pair, we assign the more massive member as the primary and the less massive as the neighbor (secondary). To enable a robust analysis of anisotropy, two final, stringent criteria are applied to the primary galaxies. First, to ensure a reliable determination of its major and minor axes, we require G1 to be viewed relatively edge-on, with an axis ratio of $b/a < 0.5$. Second, we re-enforce the requirement of high-quality Balmer lines (S/N $>$ 5 in both H$\alpha$ and H$\beta$) for G1. This requirement serves a threefold purpose: it acts as a physical filter to prioritize primary galaxies with vigorous star formation---the main drivers of metal-enriched outflows---and it ensures the high-precision dust-extinction correction required to detect the subtle, secondary environmental metallicity signal, and it provides a robust, high-S/N reference baseline for the MZR. Third, to ensure a strictly isolated hierarchical structure, we require that any galaxy identified as G2 does not simultaneously serve as G1 for another satellite. In any multi-galaxy system, only the most massive galaxy is designated as the primary. Finally, this selection yields a final sample of 1,433 galaxy pairs, comprising 2,830 unique galaxies. The sample is dominated by isolated binary systems, with only 36 primary galaxies hosting more than one neighbor, ensuring that calculating signal contribution from multiple satellites is negligible.

\subsection*{Galaxy Orientation and Geometry}\label{sec:method_geometry}

To determine the geometric orientation of each galaxy, we utilize the measured ellipticity components, $\epsilon_1$ and $\epsilon_2$, sourced directly from the DESI Legacy Survey photometric catalog (available at \url{https://www.legacysurvey.org/dr10/catalogs/}). These parameters define the complex ellipticity, $\epsilon$, which is fundamentally related to the physical shape and orientation of the galaxy on the sky:
\begin{equation}
\label{eq:complex_ellip}
\epsilon = \frac{a-b}{a+b} \exp (2 i \theta_0) = \epsilon_1 + i \epsilon_2
\end{equation}
where $a$ and $b$ are the semi-major and semi-minor axes, respectively, and $\theta_0$ is the orientation of each galaxy.

From the complex ellipticity components, we derive the physical axis ratio and the azimuthal angle of the major axis using standard relations:
\begin{equation}
\frac{b}{a} = \frac{1-|\epsilon|}{1+|\epsilon|} ;
\end{equation}
\begin{equation}
\label{eq:theta_calc}
\theta_0 = \frac{1}{2} \arctan \left( \frac{\epsilon_2}{\epsilon_1} \right).
\end{equation}
The angle $\theta_0$ is calculated relative to the image coordinate system and then transformed to the azimuthal angle $\theta$ such that it represents the angle between the primary galaxy's major axis and the vector connecting G1 to its neighbor (G2). Given the physical expectation that outflows exhibit radial symmetry about the disk, all angles are subsequently folded into the range $[0, \pi/2]$ radians, where $\theta = \pi/2$ corresponds to alignment with the minor axis (i.e., perpendicular to the disk).

\subsection*{Gas-phase Metallicity Measurement}\label{sec:method_metallicity}

For each galaxy, we determine the gas-phase metallicity ($12+\log(\text{O/H})$) using the \texttt{N2S2H$\alpha$} calibrator from Dopita et al. (2016) \citep{Dopita2016} as our fiducial diagnostic. This calibrator is advantageous as it is largely insensitive to reddening and the ionization parameter \citep[e.g.,][]{Zhang2017, Hwang2019}. Moreover, its near-linear relation with direct, electron-temperature-based metallicity measurements makes it particularly well-suited for studies of metal distribution \citep{Easeman2024}. We then parameterize the MZR of the parent sample with the following linear fit, which serves as the baseline for our metallicity offset calculations:
\begin{equation}
\label{eq:mzr_desi}
12 + \log(\text{O/H}) = (0.348 \pm 0.001) \times \log(M_*/\mathrm{M_\odot}) + (4.692 \pm 0.005)
\end{equation}
where $12+\log(\mathrm{O/H})$ is the gas-phase oxygen abundance, $M_*$ is the stellar mass, and $\mathrm{M}_\odot$ denotes the solar mass. The formal uncertainties on the fit parameters are negligible due to the large sample size. As shown in Fig.~\ref{MZR}, while the relation visually appears slightly shallower than the high-density ridge (mode) due to the asymmetric scatter of the MZR towards lower metallicities, the linear fit captures the mean trend required for calculating relative offsets.

To ensure our results are not dependent on this specific choice, we systematically test the robustness of our findings using four additional widely-used metallicity calibrators in addition to our fiducial N2S2H$\alpha$ diagnostic. These include the S-calibration (\texttt{Scal}) \citep{Pilyugin2016}, the \texttt{N2O2} diagnostic \citep{Dopita2013, Zhang2017}, \texttt{O3N2} diagnostic \citep{Marino2013}, and $R_{23}$ diagnostic \citep{Curti2017}. To resolve the double-valued nature of the $R_{23}$ diagnostic, we utilize the [N~II]/H$\alpha$ ratio to break the degeneracy between the upper and lower metallicity branches, noting that the vast majority of our DESI galaxies naturally reside on the upper branch. As shown in \ref{alterZ}, the positive trend between the neighbor's metallicity offset and its azimuthal angle in closer pairs ($15<D<60$~kpc) found using our fiducial N2S2H$\alpha$ diagnostic is broadly consistent across all four additional metallicity diagnostics. While the specific amplitude and signal-to-noise ratio vary---a reflection of the well-known systematic differences and varying sensitivities among these distinct calibrators---the key qualitative feature, namely the enrichment excess along the minor axis relative to the major axis, is recovered in most cases. This cross-calibrator consistency suggests that the detected anisotropic metal excess is a robust physical phenomenon rather than a systematic artifact inherent to any single metallicity diagnostic.

\subsection*{Statistical Analysis of Azimuthal Dependence}\label{sec:method_statistics}

To quantify the angular dependence and mitigate the effect of individual outliers in our limited sample, we utilize a running median technique for the azimuthal trends. The $1\sigma$ confidence intervals of the running medians are estimated using bootstrap resampling with $1,000$ iterations. To evaluate the statistical significance and the amplitude of the anisotropic signal, we perform a weighted linear least-squares fit to the running median points, weighted by the inverse variance of their bootstrap errors. From this fit, we extract the slope ($k$), the extrapolated total metallicity enhancement ($\Delta$) with corresponding 1$\sigma$ uncertainties, the Pearson correlation coefficient ($r$), and the $p$-value for each bin.

To evaluate whether the observed positive correlation between $\theta$ and the ($\Delta \log(\text{O/H})_{\text{G2}}$) could arise from random fluctuations or artifact of the binning procedure, we also perform Monte Carlo permutation tests. We evaluate all twelve physical bins---covering both DESI and TNG samples across various separations (D1--D3) and primary masses (M1--M3)---to ensure statistical transparency (\ref{fig:MCtest}). For each bin, we keep the observed metallicity offsets fixed but randomly shuffled $\theta$ for $10,000$ times to construct null distributions.

For each of the $10,000$ randomized datasets, we re-evaluated the slope $k$ using two independent approaches presented together in each panel of \ref{fig:MCtest}: (i) by fitting the running median points (blue, the same method as Fig.~\ref{theta-dZ}), and (ii) by a direct linear fit to the raw unbinned scatter data (orange). For metallicity excess trend, CL$ = 1 - N_{\ge k_{\text{obs}}} / 10000$, where $N_{\ge k_{\text{obs}}}$ is the number of randomized realizations that yielded a slope equal to or steeper than our true observed slope $k_{\text{obs}}$.

As shown in \ref{fig:MCtest}, the observed slopes for the key signal bins systematically reside in the right tail of the random distributions. The two independent fitting methods yield roughly consistent results across all panels. For the DESI sample, the statistical evidence for the anisotropic metal excess is most robust at intermediate separations (Bin D2), where the signal is significant (CL $> 98\%$). In other regimes, the detection appears more suggestive or marginal: in the innermost environment (Bin D1), the CL is $> 92\%$, while for the most massive primary galaxies (Bin M3), the CL ranges from $> 89\%$ to $>95\%$ depending on the fitting method. In contrast, the control samples at large separations (Bin D3) show slopes that fall within the central bulk of the null distributions (CL approximately $50\%$), suggesting the lack of a detectable feedback signal at these scales.

We acknowledge that the points in a running median are not strictly independent due to the overlapping windows, which can potentially lead to an overestimation of significance in traditional parametric tests. However, our permutation approach naturally accounts for this dependency: by applying the identical running-median and fitting procedure to each randomized dataset, the resulting null distribution of slopes ($k_{\text{rand}}$) inherently incorporates the effects of covariance between adjacent bins. The CL we report are thus derived empirically and remain robust against the statistical dependencies introduced by the smoothing procedure.

\subsection*{Galaxy Pairs in the IllustrisTNG Simulation}\label{sec:method_tng}

To provide a theoretical counterpart and compare simulation results with our observational findings, we construct a mock galaxy pair catalog from the IllustrisTNG cosmological simulation. We utilize the TNG100-1\cite{Nelson2018,Pillepich2018} simulation at the $z=0.3$ snapshot, chosen to approximate the median redshift of our DESI sample.

Our DESI sample is inherently biased toward star-forming galaxies due to the emission-line selection criteria. To mimic this, we first construct a parent sample of star-forming galaxies from the simulation, as shown in \ref{fig:TNG-pair}\textbf{a}. We select galaxies with stellar masses $\log(M_*/\mathrm{M_\odot})>8.0$ that lie above a line 1 dex below the star-forming main sequence (the SFMS$-1$dex line), given by:
\begin{equation}
\label{eq:sfms_cut}
\log(\text{SFR}) > 0.86 \times \log(M_*/\mathrm{M_\odot}) - 8.5 - 1.0
\end{equation}
where $\mathrm{SFR}$ is the star formation rate. We further apply a cut (boundary 1) to exclude massive, quenched galaxies that are unlikely to be present in our emission-line sample, defined as:
\begin{equation}
\label{eq:boundary1}
\log(\text{SFR}) > 3.3 \times \log(M_*/\mathrm{M_\odot}) - 35.
\end{equation}
Additionally, to ensure our sample consists of physically realistic galaxies, we remove systems with artificially small sizes using a cut on the mass-size relation (boundary 2 in \ref{fig:TNG-pair}\textbf{b}), defined by:
\begin{equation}
\label{eq:boundary2}
\log(R_\text{e}/\text{kpc}) > 0.05 \times \log(M_*/\mathrm{M_\odot}) - 0.3
\end{equation} 
where $R_\text{e}$ is the effective radius. This procedure yields a parent sample of 34,187 star-forming galaxies. This parent sample follows a tight MZR, which serves as our baseline for calculating metallicity offsets. We parameterize this relation with the following linear fit:
\begin{equation}
\label{eq:mzr_tng}
12 + \log(\text{O/H}) = (0.340\pm0.001) \times \log(M_*/\mathrm{M_\odot}) + (5.573\pm0.009).
\end{equation}

To create a mock observational pair catalog, first, we project the simulation's three-dimensional positions, velocities, and angular momenta of each galaxy onto the $x$-, $y$-, and $z$-plane, respectively. Second, we identify galaxy pairs using criteria identical to those applied to the DESI data: a projected pair separation of $15 < D < 100$~kpc and a projected velocity difference of $|\Delta v| < 500\ \text{km s}^{-1}$. We designate the more massive galaxy in each pair as the primary (G1). Third, to match the observational edge-on selection ($b/a < 0.5$), we require the primary galaxy's angular momentum vector $\mathbf{L}$ to have a line-of-sight component satisfying $|L_z| / |\mathbf{L}| < 0.5$. 
Finally, we apply the same rigorous hierarchical cleaning: in any multi-galaxy system, only the most massive galaxy is designated as the primary, and a galaxy identified as a neighbor (G2) is prohibited from serving as a primary for another satellite. 
This procedure yields a final sample of 9,897 galaxy pairs for three projections together in the mock data. Within each pair, the orientation between G1 and its neighbor G2 ($\theta$) is calculated using the projected angular momentum of G1 and the projected vector connecting G1 to G2, providing a direct analog to the observational measurements. The MZR and geometric parameters distributions are shown in \ref{fig:TNG-pair}\textbf{c,d,e}.

We note that some discrepancies exist in the sample distributions between DESI and TNG (e.g., the contour of the MZR in \ref{fig:TNG-pair}\textbf{c} vs. Fig.~\ref{MZR}\textbf{a}, and the pair separation distribution in \ref{fig:TNG-pair}\textbf{e} vs. Fig.~\ref{MZR}\textbf{c}). These differences are expected, as simulated metallicities are subject to sub-grid physics and nucleosynthetic yield assumptions, while the observed pair distribution is shaped by survey-specific selection effects such as fiber assignment constraints. However, since our analysis focuses on the relative metallicity offset ($\Delta\log(\mathrm{O/H})$) and the angular gradient, these distributional differences do not affect our fundamental conclusion regarding the anisotropic metal excess.

We also note that previous analysis of TNG50 by \cite{Weng2024} specifically predicted that gas outflowing along the minor axis is approximately $0.2$--$0.5$ dex more metal-rich than inflowing gas. While observational confirmation using background quasar absorption lines has proven difficult---likely due to ionization modeling uncertainties and limited sample statistics~\cite{Weng2023}---our approach using galaxy pairs provides a complementary test of these predictions by integrating the enrichment history recorded in the companion's interstellar medium.

\subsection*{Velocity and Metallicity Map around Galaxies in IllustrisTNG}\label{sec:method_map}

Having established a statistical correlation between neighbor metallicity and orientation in both DESI and the TNG simulation, we now turn to the simulation to visualize the underlying plausible physical mechanism. \ref{fig:TNG-map} presents a composite, stacked view of the gas environment around a sample of typical disk galaxies with stellar mass $\log(M_*/\mathrm{M_\odot}) = 9.9-10.1$ from the TNG100 simulation at $z = 0.3$ (matching the median redshift of our observational sample), oriented edge-on to show the gas flows relative to the disk.

\ref{fig:TNG-map}\textbf{a}, showing gas velocity streamlines, provides evidence of anisotropic galactic outflows. A powerful, wide-angle bipolar outflow is launched perpendicular to the galactic disk (along the $z$-axis), carrying gas out into the CGM and IGM at velocities of 50-80 km s$^{-1}$. The flow pattern along the major axis ($x$-axis) is significantly different, dominated by slower, co-rotating gas or inflowing gas from the cosmic web.

\ref{fig:TNG-map}\textbf{b} reveals the critical chemical consequence of this anisotropic flow. The gas-phase metallicity map shows a striking biconical structure of highly enriched gas that aligns with the bipolar outflow cones. This metal-rich material, originating from stellar feedback (e.g., supernova) within the primary galaxy, is transported into the halo, creating a chemically enriched environment along the galaxy's minor axis. In contrast, the regions along the major axis are filled with significantly more metal-poor gas.

This visualization provides a physical scenario consistent with our central hypothesis. A neighboring galaxy (G2) that lies along the minor axis of its primary (G1) is highly likely to be situated within one of these metal-enriched cones, where it can accrete this processed material, thus elevating its own gas-phase metallicity. Conversely, a neighbor along the major axis would encounter significantly less enriched gas. Although our pair analysis probes physical distances that are a subset of the large region shown here, this figure shows the persistent, large-scale nature of the enrichment process. This provides a clear, plausible link between the bipolar outflows powered by stellar feedback and the anisotropic metallicity patterns observed in our galaxy pair samples. While our current analysis of the CGM reservoir provides a spatial foundation for this interpretation, we emphasize that a definitive causal link---specifically the direct transport of gas from G1 into G2’s ISM---remains to be unambiguously established through future gas-tracing studies.

\subsection*{Sample Balance and Selection Effects}\label{sec:method_balance}

To investigate whether the anisotropic metallicity signal might be influenced by underlying selection biases, we examine the balance of physical properties of G2 between the major-axis ($0 < \theta < \pi/4$) and minor-axis ($\pi/4 < \theta < \pi/2$) subsamples. For the DESI sample, we perform KS tests on the stellar mass ($\log(M_{\text{*,G2}}/\mathrm{M_\odot})$), primary-to-secondary mass ratio ($\log M_{\text{*,G1}}/M_{\text{*,G2}}$), star formation rate ($\log \text{SFR}_{\text{G2}}$), redshift ($z_{\text{G2}}$), projected pair separation ($D$), and fiber coverage fraction of G2 ($R_{\text{fiber}}/R_\text{e,G2}$). For the TNG simulation, we perform equivalent tests on $\log M_{\text{*,G2}}$, ($\log M_{\text{*,G1}}/M_{\text{*,G2}}$), $\log \text{SFR}_{\text{G2}}$, $D$, and the effective radius ($R_{\text{e,G2}}$).

As shown in ~\ref{fig:SampleBalance} and \ref{fig:TNG-SampleBalance}, the distributions for the major-axis and minor-axis subsamples are generally statistically indistinguishable, with most $p$-values exceeding 0.05. In the TNG sample, we observe a minor difference in $\text{SFR}_{\text{G2}}$ ($p=0.004$); however, given that the primary driver of the MZR is stellar mass and that the stellar mass distribution remains balanced ($p=0.115$), this subtle variation in SFR does not qualitatively affect our conclusions. Overall, these tests indicate that our azimuthal subsamples are well-balanced in key physical properties, reducing the probability that systematic selection biases drive the detected anisotropic metal excess.

We further investigate whether the neighbor's mass ($M_{\text{*,G2}}$) or the mass ratio ($\log M_{\text{*,G1}}/M_{\text{*,G2}}$) physically modulates the anisotropic signal. While lower-mass neighbors show systematically higher average metallicity offsets---a statistical consequence of the MZR's curvature---the positive azimuthal gradient remains a persistent feature across various mass and mass-ratio bins. Since these secondary parameters are uniformly distributed relative to the primary's orientation (\ref{fig:SampleBalance}), they are unlikely to drive the detected angular trend.

\subsection*{Azimuthal Distribution of SFR Offset}\label{sec:method_sfr}

To test whether the observed metallicity excess could be a secondary effect of interaction-induced star formation, we calculate the SFR offset for the G2. We define the SFR offset, $\Delta \log(\text{SFR}_\text{G2})$, relative to the SFMS derived from our parent sample. We then apply the identical running median and bootstrap error analysis to evaluate $\Delta \log(\text{SFR}_\text{G2})$ as a function of the azimuthal angle $\theta$. To provide a quantitative statistical baseline, we further perform the Monte Carlo permutation tests on the SFR offset angular gradient ($k_{\text{SFR}}$), mirroring the methodology used for the metallicity analysis (see \ref{fig:sfr_azimuthal}).

As discussed in the main text and shown in \ref{fig:sfr_azimuthal}, the SFR offset exhibits a distinct azimuthal distribution that may be spatially decoupled from the metallicity signal. While the metallicity excess shows a positive gradient ($k > 0$), the SFR offset displays a negative angular correlation (e.g., $k_{\text{SFR}} = -0.071 \pm 0.020$ for Bin D1, $p = 0.002$). The permutation tests show that these negative slopes reside in the far-left tails of the null distributions, yielding CL approximately 86--91\% (Bin D1) and 81--86\% (Bin D2) for a negative trend (\ref{fig:sfr_azimuthal}\textbf{b, c}). Broadly, the SFR peaks along the major axis and decreases toward the minor axis. This spatial decoupling is consistent with a scenario where in-situ star formation may not be the primary driver of the minor-axis metallicity enhancement.

\bmhead{Data availability}
All data presented in this paper are publicly available. The DESI DR1 data can be accessed at https://data.desi.lbl.gov/doc/releases/dr1/. The DESI Legacy Imaging Surveys catalog can be accessed at https://www.legacysurvey.org/dr10/catalogs/. IllustrisTNG100 data can be accessed at https://www.tng-project.org/data/downloads/TNG100-1/. The generated catalog of 1,433 galaxy pairs constructed in this study, including their coordinates, redshifts, mass, SFR, and derived gas-phase metallicities and metallicity offsets, has been deposited in the Zenodo database https://doi.org/10.5281/zenodo.21838127. Source data to generate figures are provided with this paper.

\bmhead{Code availability}
The code for data analysis written in python are publicly available on the Zenodo database 10.5281/zenodo.21822727.

\sloppy  
\bibliography{sn-bibliography}

\begin{thebibliography}{10}
\expandafter\ifx\csname url\endcsname\relax
  \def\url#1{\burl{#1}}\fi
\expandafter\ifx\csname urlprefix\endcsname\relax\def\urlprefix{URL }\fi
\providecommand{\bibinfo}[2]{#2}
\providecommand{\eprint}[2][]{\url{#2}}
\providecommand{\doi}[1]{\url{https://doi.org/#1}}
\bibcommenthead

\bibitem{White1978}
\bibinfo{author}{{White}, S.~D.~M.} \& \bibinfo{author}{{Rees}, M.~J.}
\newblock \bibinfo{title}{{Core condensation in heavy halos: a two-stage theory for galaxy formation and clustering.}}
\newblock \emph{\bibinfo{journal}{\mnras}} \textbf{\bibinfo{volume}{183}}, \bibinfo{pages}{341--358} (\bibinfo{year}{1978}).

\bibitem{White1991}
\bibinfo{author}{{White}, S. D.~M.} \& \bibinfo{author}{{Frenk}, C.~S.}
\newblock \bibinfo{title}{{Galaxy Formation through Hierarchical Clustering}}.
\newblock \emph{\bibinfo{journal}{\apj}} \textbf{\bibinfo{volume}{379}}, \bibinfo{pages}{52} (\bibinfo{year}{1991}).

\bibitem{Veilleux2005}
\bibinfo{author}{{Veilleux}, S.}, \bibinfo{author}{{Cecil}, G.} \& \bibinfo{author}{{Bland-Hawthorn}, J.}
\newblock \bibinfo{title}{{Galactic Winds}}.
\newblock \emph{\bibinfo{journal}{\araa}} \textbf{\bibinfo{volume}{43}}, \bibinfo{pages}{769--826} (\bibinfo{year}{2005}).

\bibitem{Murray2005}
\bibinfo{author}{{Murray}, N.}, \bibinfo{author}{{Quataert}, E.} \& \bibinfo{author}{{Thompson}, T.~A.}
\newblock \bibinfo{title}{{On the Maximum Luminosity of Galaxies and Their Central Black Holes: Feedback from Momentum-driven Winds}}.
\newblock \emph{\bibinfo{journal}{\apj}} \textbf{\bibinfo{volume}{618}}, \bibinfo{pages}{569--585} (\bibinfo{year}{2005}).

\bibitem{Behroozi2013}
\bibinfo{author}{{Behroozi}, P.~S.}, \bibinfo{author}{{Wechsler}, R.~H.} \& \bibinfo{author}{{Conroy}, C.}
\newblock \bibinfo{title}{{The Average Star Formation Histories of Galaxies in Dark Matter Halos from z = 0-8}}.
\newblock \emph{\bibinfo{journal}{\apj}} \textbf{\bibinfo{volume}{770}}, \bibinfo{pages}{57} (\bibinfo{year}{2013}).

\bibitem{Moster2013}
\bibinfo{author}{{Moster}, B.~P.}, \bibinfo{author}{{Naab}, T.} \& \bibinfo{author}{{White}, S. D.~M.}
\newblock \bibinfo{title}{{Galactic star formation and accretion histories from matching galaxies to dark matter haloes}}.
\newblock \emph{\bibinfo{journal}{\mnras}} \textbf{\bibinfo{volume}{428}}, \bibinfo{pages}{3121--3138} (\bibinfo{year}{2013}).

\bibitem{Hopkins2014}
\bibinfo{author}{{Hopkins}, P.~F.} \emph{et~al.}
\newblock \bibinfo{title}{{Galaxies on FIRE (Feedback In Realistic Environments): stellar feedback explains cosmologically inefficient star formation}}.
\newblock \emph{\bibinfo{journal}{\mnras}} \textbf{\bibinfo{volume}{445}}, \bibinfo{pages}{581--603} (\bibinfo{year}{2014}).

\bibitem{Somerville2015}
\bibinfo{author}{{Somerville}, R.~S.} \& \bibinfo{author}{{Dav{\'e}}, R.}
\newblock \bibinfo{title}{{Physical Models of Galaxy Formation in a Cosmological Framework}}.
\newblock \emph{\bibinfo{journal}{\araa}} \textbf{\bibinfo{volume}{53}}, \bibinfo{pages}{51--113} (\bibinfo{year}{2015}).

\bibitem{Heckman2017}
\bibinfo{author}{{Heckman}, T.}, \bibinfo{author}{{Borthakur}, S.}, \bibinfo{author}{{Wild}, V.}, \bibinfo{author}{{Schiminovich}, D.} \& \bibinfo{author}{{Bordoloi}, R.}
\newblock \bibinfo{title}{{COS-burst: Observations of the Impact of Starburst-driven Winds on the Properties of the Circum-galactic Medium}}.
\newblock \emph{\bibinfo{journal}{\apj}} \textbf{\bibinfo{volume}{846}}, \bibinfo{pages}{151} (\bibinfo{year}{2017}).

\bibitem{Lyu2023}
\bibinfo{author}{{Lyu}, C.} \emph{et~al.}
\newblock \bibinfo{title}{{From Halos to Galaxies. VII. The Connections between Stellar Mass Growth History, Quenching History, and Halo Assembly History for Central Galaxies}}.
\newblock \emph{\bibinfo{journal}{\apj}} \textbf{\bibinfo{volume}{959}}, \bibinfo{pages}{5} (\bibinfo{year}{2023}).

\bibitem{Mo2024}
\bibinfo{author}{{Mo}, H.}, \bibinfo{author}{{Chen}, Y.} \& \bibinfo{author}{{Wang}, H.}
\newblock \bibinfo{title}{{A two-phase model of galaxy formation: I. The growth of galaxies and supermassive black holes}}.
\newblock \emph{\bibinfo{journal}{\mnras}} \textbf{\bibinfo{volume}{532}}, \bibinfo{pages}{3808--3838} (\bibinfo{year}{2024}).

\bibitem{Li2025}
\bibinfo{author}{{Li}, H.}, \bibinfo{author}{{Chen}, Y.}, \bibinfo{author}{{Wang}, H.} \& \bibinfo{author}{{Mo}, H.}
\newblock \bibinfo{title}{{Physical processes behind the co-evolution of haloes, galaxies, and supermassive black holes in the IllustrisTNG simulation}}.
\newblock \emph{\bibinfo{journal}{\mnras}} \textbf{\bibinfo{volume}{543}}, \bibinfo{pages}{1878--1898} (\bibinfo{year}{2025}).

\bibitem{Zhao2025}
\bibinfo{author}{{Zhao}, D.} \emph{et~al.}
\newblock \bibinfo{title}{{From Halos to Galaxies. VI. Improved Halo Mass Estimation for SDSS Groups and Measurement of the Halo Mass Function}}.
\newblock \emph{\bibinfo{journal}{\apj}} \textbf{\bibinfo{volume}{979}}, \bibinfo{pages}{42} (\bibinfo{year}{2025}).

\bibitem{WangK2025}
\bibinfo{author}{{Wang}, K.} \& \bibinfo{author}{{Peng}, Y.}
\newblock \bibinfo{title}{{Testing Galaxy Formation Models with the Stellar Mass─Halo Mass Relations for Star-forming and Quiescent Galaxies}}.
\newblock \emph{\bibinfo{journal}{\apj}} \textbf{\bibinfo{volume}{980}}, \bibinfo{pages}{233} (\bibinfo{year}{2025}).

\bibitem{WangK2026a}
\bibinfo{author}{{Wang}, K.} \emph{et~al.}
\newblock \bibinfo{title}{{Gravitational potential drives the concentration dependence of the stellar mass─halo mass relation}}.
\newblock \emph{\bibinfo{journal}{\mnras}} \textbf{\bibinfo{volume}{546}}, \bibinfo{pages}{stag110} (\bibinfo{year}{2026}).

\bibitem{Oppenheimer2006}
\bibinfo{author}{{Oppenheimer}, B.~D.} \& \bibinfo{author}{{Dav{\'e}}, R.}
\newblock \bibinfo{title}{{Cosmological simulations of intergalactic medium enrichment from galactic outflows}}.
\newblock \emph{\bibinfo{journal}{\mnras}} \textbf{\bibinfo{volume}{373}}, \bibinfo{pages}{1265--1292} (\bibinfo{year}{2006}).

\bibitem{Tumlinson2017}
\bibinfo{author}{{Tumlinson}, J.}, \bibinfo{author}{{Peeples}, M.~S.} \& \bibinfo{author}{{Werk}, J.~K.}
\newblock \bibinfo{title}{{The Circumgalactic Medium}}.
\newblock \emph{\bibinfo{journal}{\araa}} \textbf{\bibinfo{volume}{55}}, \bibinfo{pages}{389--432} (\bibinfo{year}{2017}).

\bibitem{Peroux2020}
\bibinfo{author}{{P{\'e}roux}, C.} \& \bibinfo{author}{{Howk}, J.~C.}
\newblock \bibinfo{title}{{The Cosmic Baryon and Metal Cycles}}.
\newblock \emph{\bibinfo{journal}{\araa}} \textbf{\bibinfo{volume}{58}}, \bibinfo{pages}{363--406} (\bibinfo{year}{2020}).

\bibitem{Veilleux2020}
\bibinfo{author}{{Veilleux}, S.}, \bibinfo{author}{{Maiolino}, R.}, \bibinfo{author}{{Bolatto}, A.~D.} \& \bibinfo{author}{{Aalto}, S.}
\newblock \bibinfo{title}{{Cool outflows in galaxies and their implications}}.
\newblock \emph{\bibinfo{journal}{\aapr}} \textbf{\bibinfo{volume}{28}}, \bibinfo{pages}{2} (\bibinfo{year}{2020}).

\bibitem{Wang2022a}
\bibinfo{author}{{Wang}, E.} \& \bibinfo{author}{{Lilly}, S.~J.}
\newblock \bibinfo{title}{{Gas-phase Metallicity Profiles of Star-forming Galaxies in the Modified Accretion Disk Framework}}.
\newblock \emph{\bibinfo{journal}{\apj}} \textbf{\bibinfo{volume}{929}}, \bibinfo{pages}{95} (\bibinfo{year}{2022}).

\bibitem{Wang2022b}
\bibinfo{author}{{Wang}, E.} \& \bibinfo{author}{{Lilly}, S.~J.}
\newblock \bibinfo{title}{{The Origin of Exponential Star-forming Disks}}.
\newblock \emph{\bibinfo{journal}{\apj}} \textbf{\bibinfo{volume}{927}}, \bibinfo{pages}{217} (\bibinfo{year}{2022}).

\bibitem{Faucher-Giguere2023}
\bibinfo{author}{{Faucher-Gigu{\`e}re}, C.-A.} \& \bibinfo{author}{{Oh}, S.~P.}
\newblock \bibinfo{title}{{Key Physical Processes in the Circumgalactic Medium}}.
\newblock \emph{\bibinfo{journal}{\araa}} \textbf{\bibinfo{volume}{61}}, \bibinfo{pages}{131--195} (\bibinfo{year}{2023}).

\bibitem{Ma2024}
\bibinfo{author}{{Ma}, C.} \emph{et~al.}
\newblock \bibinfo{title}{{Revisiting the Fundamental Metallicity Relation with Observation and Simulation}}.
\newblock \emph{\bibinfo{journal}{\apjl}} \textbf{\bibinfo{volume}{971}}, \bibinfo{pages}{L14} (\bibinfo{year}{2024}).

\bibitem{ChenYY2025}
\bibinfo{author}{{Chen}, Y.}, \bibinfo{author}{{Mo}, H.} \& \bibinfo{author}{{Wang}, H.}
\newblock \bibinfo{title}{{A two-phase model of galaxy formation: IV. Seeding and growing supermassive black holes in dark matter halos}}.
\newblock \emph{\bibinfo{journal}{arXiv e-prints}} \bibinfo{pages}{arXiv:2509.03283} (\bibinfo{year}{2025}).

\bibitem{Yu2026}
\bibinfo{author}{{Yu}, H.} \emph{et~al.}
\newblock \bibinfo{title}{{Stellar feedback drives the baryon deficiency in low-mass galaxies}}.
\newblock \emph{\bibinfo{journal}{Science Advances}} \textbf{\bibinfo{volume}{12}}, \bibinfo{pages}{4.7506} (\bibinfo{year}{2026}).

\bibitem{Lyu2026}
\bibinfo{author}{{Lyu}, C.} \emph{et~al.}
\newblock \bibinfo{title}{{First Statistical Detection of Mg II-traced Cool Gas Outflows with JWST toward Cosmic Dawn}}.
\newblock \emph{\bibinfo{journal}{\apjl}} \textbf{\bibinfo{volume}{1000}}, \bibinfo{pages}{L3} (\bibinfo{year}{2026}).

\bibitem{WangK2026b}
\bibinfo{author}{{Wang}, K.}
\newblock \bibinfo{title}{{The origin of the galaxy size─stellar metallicity relation ─ I. A semi-analytical perspective}}.
\newblock \emph{\bibinfo{journal}{\mnras}} \textbf{\bibinfo{volume}{545}}, \bibinfo{pages}{staf2113} (\bibinfo{year}{2026}).

\bibitem{Heckman1990}
\bibinfo{author}{{Heckman}, T.~M.}, \bibinfo{author}{{Armus}, L.} \& \bibinfo{author}{{Miley}, G.~K.}
\newblock \bibinfo{title}{{On the Nature and Implications of Starburst-driven Galactic Superwinds}}.
\newblock \emph{\bibinfo{journal}{\apjs}} \textbf{\bibinfo{volume}{74}}, \bibinfo{pages}{833} (\bibinfo{year}{1990}).

\bibitem{Muratov2015}
\bibinfo{author}{{Muratov}, A.~L.} \emph{et~al.}
\newblock \bibinfo{title}{{Gusty, gaseous flows of FIRE: galactic winds in cosmological simulations with explicit stellar feedback}}.
\newblock \emph{\bibinfo{journal}{\mnras}} \textbf{\bibinfo{volume}{454}}, \bibinfo{pages}{2691--2713} (\bibinfo{year}{2015}).

\bibitem{Nelson2019}
\bibinfo{author}{{Nelson}, D.} \emph{et~al.}
\newblock \bibinfo{title}{{First results from the TNG50 simulation: galactic outflows driven by supernovae and black hole feedback}}.
\newblock \emph{\bibinfo{journal}{\mnras}} \textbf{\bibinfo{volume}{490}}, \bibinfo{pages}{3234--3261} (\bibinfo{year}{2019}).

\bibitem{Guo2023}
\bibinfo{author}{{Guo}, Y.} \emph{et~al.}
\newblock \bibinfo{title}{{Bipolar outflows out to 10 kpc for massive galaxies at redshift z {\ensuremath{\approx}} 1}}.
\newblock \emph{\bibinfo{journal}{\nat}} \textbf{\bibinfo{volume}{624}}, \bibinfo{pages}{53--56} (\bibinfo{year}{2023}).

\bibitem{Chen2025}
\bibinfo{author}{{Chen}, Z.} \emph{et~al.}
\newblock \bibinfo{title}{{The Circumgalactic Medium Traced by Mg II Absorption with DESI: Dependence on Galaxy Stellar Mass, Star Formation Rate, and Azimuthal Angle}}.
\newblock \emph{\bibinfo{journal}{\apj}} \textbf{\bibinfo{volume}{981}}, \bibinfo{pages}{81} (\bibinfo{year}{2025}).

\bibitem{Zhang2024}
\bibinfo{author}{{Zhang}, H.}, \bibinfo{author}{{Li}, M.} \& \bibinfo{author}{{Zaritsky}, D.}
\newblock \bibinfo{title}{{The Anisotropic Circumgalactic Medium of Sub-L* Galaxies}}.
\newblock \emph{\bibinfo{journal}{\apj}} \textbf{\bibinfo{volume}{974}}, \bibinfo{pages}{148} (\bibinfo{year}{2024}).

\bibitem{Wendt2021}
\bibinfo{author}{{Wendt}, M.}, \bibinfo{author}{{Bouch{\'e}}, N.~F.}, \bibinfo{author}{{Zabl}, J.}, \bibinfo{author}{{Schroetter}, I.} \& \bibinfo{author}{{Muzahid}, S.}
\newblock \bibinfo{title}{{MusE GAs FLOw and Wind V. The dust/metallicity-anisotropy of the circum-galactic medium}}.
\newblock \emph{\bibinfo{journal}{\mnras}} \textbf{\bibinfo{volume}{502}}, \bibinfo{pages}{3733--3745} (\bibinfo{year}{2021}).

\bibitem{Tremonti2004}
\bibinfo{author}{{Tremonti}, C.~A.} \emph{et~al.}
\newblock \bibinfo{title}{{The Origin of the Mass-Metallicity Relation: Insights from 53,000 Star-forming Galaxies in the Sloan Digital Sky Survey}}.
\newblock \emph{\bibinfo{journal}{\apj}} \textbf{\bibinfo{volume}{613}}, \bibinfo{pages}{898--913} (\bibinfo{year}{2004}).

\bibitem{Ellison2008}
\bibinfo{author}{{Ellison}, S.~L.}, \bibinfo{author}{{Patton}, D.~R.}, \bibinfo{author}{{Simard}, L.} \& \bibinfo{author}{{McConnachie}, A.~W.}
\newblock \bibinfo{title}{{Galaxy Pairs in the Sloan Digital Sky Survey. I. Star Formation, Active Galactic Nucleus Fraction, and the Mass-Metallicity Relation}}.
\newblock \emph{\bibinfo{journal}{\aj}} \textbf{\bibinfo{volume}{135}}, \bibinfo{pages}{1877--1899} (\bibinfo{year}{2008}).

\bibitem{Ellison2009}
\bibinfo{author}{{Ellison}, S.~L.} \emph{et~al.}
\newblock \bibinfo{title}{{The mass-metallicity relation in galaxy clusters: the relative importance of cluster membership versus local environment}}.
\newblock \emph{\bibinfo{journal}{\mnras}} \textbf{\bibinfo{volume}{396}}, \bibinfo{pages}{1257--1272} (\bibinfo{year}{2009}).

\bibitem{Mannucci2010}
\bibinfo{author}{{Mannucci}, F.}, \bibinfo{author}{{Cresci}, G.}, \bibinfo{author}{{Maiolino}, R.}, \bibinfo{author}{{Marconi}, A.} \& \bibinfo{author}{{Gnerucci}, A.}
\newblock \bibinfo{title}{{A fundamental relation between mass, star formation rate and metallicity in local and high-redshift galaxies}}.
\newblock \emph{\bibinfo{journal}{\mnras}} \textbf{\bibinfo{volume}{408}}, \bibinfo{pages}{2115--2127} (\bibinfo{year}{2010}).

\bibitem{Cooper2008}
\bibinfo{author}{{Cooper}, M.~C.}, \bibinfo{author}{{Tremonti}, C.~A.}, \bibinfo{author}{{Newman}, J.~A.} \& \bibinfo{author}{{Zabludoff}, A.~I.}
\newblock \bibinfo{title}{{The role of environment in the mass-metallicity relation}}.
\newblock \emph{\bibinfo{journal}{\mnras}} \textbf{\bibinfo{volume}{390}}, \bibinfo{pages}{245--256} (\bibinfo{year}{2008}).

\bibitem{Pasquali2010}
\bibinfo{author}{{Pasquali}, A.} \emph{et~al.}
\newblock \bibinfo{title}{{Ages and metallicities of central and satellite galaxies: implications for galaxy formation and evolution}}.
\newblock \emph{\bibinfo{journal}{\mnras}} \textbf{\bibinfo{volume}{407}}, \bibinfo{pages}{937--954} (\bibinfo{year}{2010}).

\bibitem{Scudder2012}
\bibinfo{author}{{Scudder}, J.~M.}, \bibinfo{author}{{Ellison}, S.~L.}, \bibinfo{author}{{Torrey}, P.}, \bibinfo{author}{{Patton}, D.~R.} \& \bibinfo{author}{{Mendel}, J.~T.}
\newblock \bibinfo{title}{{Galaxy pairs in the Sloan Digital Sky Survey - V. Tracing changes in star formation rate and metallicity out to separations of 80 kpc}}.
\newblock \emph{\bibinfo{journal}{\mnras}} \textbf{\bibinfo{volume}{426}}, \bibinfo{pages}{549--565} (\bibinfo{year}{2012}).

\bibitem{Peng2014}
\bibinfo{author}{{Peng}, Y.-j.} \& \bibinfo{author}{{Maiolino}, R.}
\newblock \bibinfo{title}{{The dependence of the galaxy mass-metallicity relation on environment and the implied metallicity of the IGM}}.
\newblock \emph{\bibinfo{journal}{\mnras}} \textbf{\bibinfo{volume}{438}}, \bibinfo{pages}{262--270} (\bibinfo{year}{2014}).

\bibitem{WangK2023}
\bibinfo{author}{{Wang}, K.}, \bibinfo{author}{{Wang}, X.} \& \bibinfo{author}{{Chen}, Y.}
\newblock \bibinfo{title}{{Environmental Dependence of the Mass-Metallicity Relation in Cosmological Hydrodynamical Simulations}}.
\newblock \emph{\bibinfo{journal}{\apj}} \textbf{\bibinfo{volume}{951}}, \bibinfo{pages}{66} (\bibinfo{year}{2023}).

\bibitem{Levi2013}
\bibinfo{author}{{Levi}, M.} \emph{et~al.}
\newblock \bibinfo{title}{{The DESI Experiment, a whitepaper for Snowmass 2013}}.
\newblock \emph{\bibinfo{journal}{arXiv e-prints}} \bibinfo{pages}{arXiv:1308.0847} (\bibinfo{year}{2013}).

\bibitem{DESICollaboration2022}
\bibinfo{author}{{DESI Collaboration}} \emph{et~al.}
\newblock \bibinfo{title}{{Overview of the Instrumentation for the Dark Energy Spectroscopic Instrument}}.
\newblock \emph{\bibinfo{journal}{\aj}} \textbf{\bibinfo{volume}{164}}, \bibinfo{pages}{207} (\bibinfo{year}{2022}).

\bibitem{Levi2019}
\bibinfo{author}{{Levi}, M.} \emph{et~al.}
\newblock \emph{\bibinfo{title}{{The Dark Energy Spectroscopic Instrument (DESI)}}}, Vol.~\bibinfo{volume}{51}, \bibinfo{pages}{57} (\bibinfo{year}{2019}).
\newblock \bibinfo{eprint}{{\href{https://arxiv.org/abs/1907.10688}{{arXiv:1907.10688}}}}.

\bibitem{Silber2023}
\bibinfo{author}{{Silber}, J.~H.} \emph{et~al.}
\newblock \bibinfo{title}{{The Robotic Multiobject Focal Plane System of the Dark Energy Spectroscopic Instrument (DESI)}}.
\newblock \emph{\bibinfo{journal}{\aj}} \textbf{\bibinfo{volume}{165}}, \bibinfo{pages}{9} (\bibinfo{year}{2023}).

\bibitem{Miller2024}
\bibinfo{author}{{Miller}, T.~N.} \emph{et~al.}
\newblock \bibinfo{title}{{The Optical Corrector for the Dark Energy Spectroscopic Instrument}}.
\newblock \emph{\bibinfo{journal}{\aj}} \textbf{\bibinfo{volume}{168}}, \bibinfo{pages}{95} (\bibinfo{year}{2024}).

\bibitem{DESICollaboration2024b}
\bibinfo{author}{{DESI Collaboration}} \emph{et~al.}
\newblock \bibinfo{title}{{The Early Data Release of the Dark Energy Spectroscopic Instrument}}.
\newblock \emph{\bibinfo{journal}{\aj}} \textbf{\bibinfo{volume}{168}}, \bibinfo{pages}{58} (\bibinfo{year}{2024}).

\bibitem{Dopita2016}
\bibinfo{author}{{Dopita}, M.~A.}, \bibinfo{author}{{Kewley}, L.~J.}, \bibinfo{author}{{Sutherland}, R.~S.} \& \bibinfo{author}{{Nicholls}, D.~C.}
\newblock \bibinfo{title}{{Chemical abundances in high-redshift galaxies: a powerful new emission line diagnostic}}.
\newblock \emph{\bibinfo{journal}{\apss}} \textbf{\bibinfo{volume}{361}}, \bibinfo{pages}{61} (\bibinfo{year}{2016}).

\bibitem{Zhang2017}
\bibinfo{author}{{Zhang}, K.} \emph{et~al.}
\newblock \bibinfo{title}{{SDSS-IV MaNGA: the impact of diffuse ionized gas on emission-line ratios, interpretation of diagnostic diagrams and gas metallicity measurements}}.
\newblock \emph{\bibinfo{journal}{\mnras}} \textbf{\bibinfo{volume}{466}}, \bibinfo{pages}{3217--3243} (\bibinfo{year}{2017}).

\bibitem{Hwang2019}
\bibinfo{author}{{Hwang}, H.-C.} \emph{et~al.}
\newblock \bibinfo{title}{{Anomalously Low-metallicity Regions in MaNGA Star-forming Galaxies: Accretion Caught in Action?}}
\newblock \emph{\bibinfo{journal}{\apj}} \textbf{\bibinfo{volume}{872}}, \bibinfo{pages}{144} (\bibinfo{year}{2019}).

\bibitem{Nelson2018}
\bibinfo{author}{{Nelson}, D.} \emph{et~al.}
\newblock \bibinfo{title}{{First results from the IllustrisTNG simulations: the galaxy colour bimodality}}.
\newblock \emph{\bibinfo{journal}{\mnras}} \textbf{\bibinfo{volume}{475}}, \bibinfo{pages}{624--647} (\bibinfo{year}{2018}).

\bibitem{Pillepich2018}
\bibinfo{author}{{Pillepich}, A.} \emph{et~al.}
\newblock \bibinfo{title}{{Simulating galaxy formation with the IllustrisTNG model}}.
\newblock \emph{\bibinfo{journal}{\mnras}} \textbf{\bibinfo{volume}{473}}, \bibinfo{pages}{4077--4106} (\bibinfo{year}{2018}).

\bibitem{Peroux2020b}
\bibinfo{author}{{P{\'e}roux}, C.} \emph{et~al.}
\newblock \bibinfo{title}{{Predictions for the angular dependence of gas mass flow rate and metallicity in the circumgalactic medium}}.
\newblock \emph{\bibinfo{journal}{\mnras}} \textbf{\bibinfo{volume}{499}}, \bibinfo{pages}{2462--2473} (\bibinfo{year}{2020}).

\bibitem{Weng2024}
\bibinfo{author}{{Weng}, S.} \emph{et~al.}
\newblock \bibinfo{title}{{The physical origins of gas in the circumgalactic medium using observationally motivated TNG50 mocks}}.
\newblock \emph{\bibinfo{journal}{\mnras}} \textbf{\bibinfo{volume}{527}}, \bibinfo{pages}{3494--3516} (\bibinfo{year}{2024}).

\bibitem{Ramesh2023}
\bibinfo{author}{{Ramesh}, R.}, \bibinfo{author}{{Nelson}, D.} \& \bibinfo{author}{{Pillepich}, A.}
\newblock \bibinfo{title}{{The circumgalactic medium of Milky Way-like galaxies in the TNG50 simulation - I: halo gas properties and the role of SMBH feedback}}.
\newblock \emph{\bibinfo{journal}{\mnras}} \textbf{\bibinfo{volume}{518}}, \bibinfo{pages}{5754--5777} (\bibinfo{year}{2023}).

\bibitem{Weng2023}
\bibinfo{author}{{Weng}, S.} \emph{et~al.}
\newblock \bibinfo{title}{{MUSE-ALMA Haloes XI: gas flows in the circumgalactic medium}}.
\newblock \emph{\bibinfo{journal}{\mnras}} \textbf{\bibinfo{volume}{523}}, \bibinfo{pages}{676--700} (\bibinfo{year}{2023}).

\bibitem{Patton2013}
\bibinfo{author}{{Patton}, D.~R.}, \bibinfo{author}{{Torrey}, P.}, \bibinfo{author}{{Ellison}, S.~L.}, \bibinfo{author}{{Mendel}, J.~T.} \& \bibinfo{author}{{Scudder}, J.~M.}
\newblock \bibinfo{title}{{Galaxy pairs in the Sloan Digital Sky Survey - VI. The orbital extent of enhanced star formation in interacting galaxies}}.
\newblock \emph{\bibinfo{journal}{\mnras}} \textbf{\bibinfo{volume}{433}}, \bibinfo{pages}{L59--L63} (\bibinfo{year}{2013}).

\bibitem{Sancisi2008}
\bibinfo{author}{{Sancisi}, R.}, \bibinfo{author}{{Fraternali}, F.}, \bibinfo{author}{{Oosterloo}, T.} \& \bibinfo{author}{{van der Hulst}, T.}
\newblock \bibinfo{title}{{Cold gas accretion in galaxies}}.
\newblock \emph{\bibinfo{journal}{\aapr}} \textbf{\bibinfo{volume}{15}}, \bibinfo{pages}{189--223} (\bibinfo{year}{2008}).

\bibitem{Rupke2010}
\bibinfo{author}{{Rupke}, D. S.~N.}, \bibinfo{author}{{Kewley}, L.~J.} \& \bibinfo{author}{{Barnes}, J.~E.}
\newblock \bibinfo{title}{{Galaxy Mergers and the Mass-Metallicity Relation: Evidence for Nuclear Metal Dilution and Flattened Gradients from Numerical Simulations}}.
\newblock \emph{\bibinfo{journal}{\apjl}} \textbf{\bibinfo{volume}{710}}, \bibinfo{pages}{L156--L160} (\bibinfo{year}{2010}).

\bibitem{Torrey2012}
\bibinfo{author}{{Torrey}, P.}, \bibinfo{author}{{Cox}, T.~J.}, \bibinfo{author}{{Kewley}, L.} \& \bibinfo{author}{{Hernquist}, L.}
\newblock \bibinfo{title}{{The Metallicity Evolution of Interacting Galaxies}}.
\newblock \emph{\bibinfo{journal}{\apj}} \textbf{\bibinfo{volume}{746}}, \bibinfo{pages}{108} (\bibinfo{year}{2012}).

\bibitem{Ellison2013}
\bibinfo{author}{{Ellison}, S.~L.}, \bibinfo{author}{{Mendel}, J.~T.}, \bibinfo{author}{{Patton}, D.~R.} \& \bibinfo{author}{{Scudder}, J.~M.}
\newblock \bibinfo{title}{{Galaxy pairs in the Sloan Digital Sky Survey - VIII. The observational properties of post-merger galaxies}}.
\newblock \emph{\bibinfo{journal}{\mnras}} \textbf{\bibinfo{volume}{435}}, \bibinfo{pages}{3627--3638} (\bibinfo{year}{2013}).

\bibitem{Zubovas2013}
\bibinfo{author}{{Zubovas}, K.}, \bibinfo{author}{{Nayakshin}, S.}, \bibinfo{author}{{King}, A.} \& \bibinfo{author}{{Wilkinson}, M.}
\newblock \bibinfo{title}{{AGN outflows trigger starbursts in gas-rich galaxies}}.
\newblock \emph{\bibinfo{journal}{\mnras}} \textbf{\bibinfo{volume}{433}}, \bibinfo{pages}{3079--3090} (\bibinfo{year}{2013}).

\bibitem{Bieri2016}
\bibinfo{author}{{Bieri}, R.}, \bibinfo{author}{{Dubois}, Y.}, \bibinfo{author}{{Silk}, J.}, \bibinfo{author}{{Mamon}, G.~A.} \& \bibinfo{author}{{Gaibler}, V.}
\newblock \bibinfo{title}{{External pressure-triggering of star formation in a disc galaxy: a template for positive feedback}}.
\newblock \emph{\bibinfo{journal}{\mnras}} \textbf{\bibinfo{volume}{455}}, \bibinfo{pages}{4166--4182} (\bibinfo{year}{2016}).

\bibitem{Heckman2017b}
\bibinfo{author}{{Heckman}, T.~M.} \& \bibinfo{author}{{Thompson}, T.~A.}
\newblock \bibinfo{title}{{Galactic Winds and the Role Played by Massive Stars}}.
\newblock \emph{\bibinfo{journal}{arXiv e-prints}} \bibinfo{pages}{arXiv:1701.09062} (\bibinfo{year}{2017}).

\bibitem{Danovich2015}
\bibinfo{author}{{Danovich}, M.}, \bibinfo{author}{{Dekel}, A.}, \bibinfo{author}{{Hahn}, O.}, \bibinfo{author}{{Ceverino}, D.} \& \bibinfo{author}{{Primack}, J.}
\newblock \bibinfo{title}{{Four phases of angular-momentum buildup in high-z galaxies: from cosmic-web streams through an extended ring to disc and bulge}}.
\newblock \emph{\bibinfo{journal}{\mnras}} \textbf{\bibinfo{volume}{449}}, \bibinfo{pages}{2087--2111} (\bibinfo{year}{2015}).

\bibitem{Stewart2017}
\bibinfo{author}{{Stewart}, K.~R.} \emph{et~al.}
\newblock \bibinfo{title}{{High Angular Momentum Halo Gas: A Feedback and Code-independent Prediction of LCDM}}.
\newblock \emph{\bibinfo{journal}{\apj}} \textbf{\bibinfo{volume}{843}}, \bibinfo{pages}{47} (\bibinfo{year}{2017}).

\bibitem{PlanckCollaboration2016}
\bibinfo{author}{{Planck Collaboration}} \emph{et~al.}
\newblock \bibinfo{title}{{Planck 2015 results. XIII. Cosmological parameters}}.
\newblock \emph{\bibinfo{journal}{\aap}} \textbf{\bibinfo{volume}{594}}, \bibinfo{pages}{A13} (\bibinfo{year}{2016}).

\bibitem{DESICollaboration2025}
\bibinfo{author}{{DESI Collaboration}} \emph{et~al.}
\newblock \bibinfo{title}{{Data Release 1 of the Dark Energy Spectroscopic Instrument}}.
\newblock \emph{\bibinfo{journal}{arXiv e-prints}} \bibinfo{pages}{arXiv:2503.14745} (\bibinfo{year}{2025}).

\bibitem{Adame2025}
\bibinfo{author}{{Adame}, A.~G.} \emph{et~al.}
\newblock \bibinfo{title}{{DESI 2024 V: Full-Shape galaxy clustering from galaxies and quasars}}.
\newblock \emph{\bibinfo{journal}{\jcap}} \textbf{\bibinfo{volume}{2025}}, \bibinfo{pages}{008} (\bibinfo{year}{2025}).

\bibitem{Adame2025a}
\bibinfo{author}{{Adame}, A.~G.} \emph{et~al.}
\newblock \bibinfo{title}{{DESI 2024 VI: cosmological constraints from the measurements of baryon acoustic oscillations}}.
\newblock \emph{\bibinfo{journal}{\jcap}} \textbf{\bibinfo{volume}{2025}}, \bibinfo{pages}{021} (\bibinfo{year}{2025}).

\bibitem{DESICollaboration2024c}
\bibinfo{author}{{DESI Collaboration}} \emph{et~al.}
\newblock \bibinfo{title}{{Validation of the Scientific Program for the Dark Energy Spectroscopic Instrument}}.
\newblock \emph{\bibinfo{journal}{\aj}} \textbf{\bibinfo{volume}{167}}, \bibinfo{pages}{62} (\bibinfo{year}{2024}).

\bibitem{Adame2025b}
\bibinfo{author}{{Adame}, A.~G.} \emph{et~al.}
\newblock \bibinfo{title}{{DESI 2024 IV: Baryon Acoustic Oscillations from the Lyman alpha forest}}.
\newblock \emph{\bibinfo{journal}{\jcap}} \textbf{\bibinfo{volume}{2025}}, \bibinfo{pages}{124} (\bibinfo{year}{2025}).

\bibitem{Hahn2023}
\bibinfo{author}{{Hahn}, C.} \emph{et~al.}
\newblock \bibinfo{title}{{The DESI Bright Galaxy Survey: Final Target Selection, Design, and Validation}}.
\newblock \emph{\bibinfo{journal}{\aj}} \textbf{\bibinfo{volume}{165}}, \bibinfo{pages}{253} (\bibinfo{year}{2023}).

\bibitem{Osterbrock2006}
\bibinfo{author}{{Osterbrock}, D.~E.} \& \bibinfo{author}{{Ferland}, G.~J.}
\newblock \emph{\bibinfo{title}{Astrophysics of Gaseous Nebulae and Active Galactic Nuclei}}  (\bibinfo{year}{2006}).

\bibitem{Cardelli1989}
\bibinfo{author}{{Cardelli}, J.~A.}, \bibinfo{author}{{Clayton}, G.~C.} \& \bibinfo{author}{{Mathis}, J.~S.}
\newblock \bibinfo{title}{{The Relationship between Infrared, Optical, and Ultraviolet Extinction}}.
\newblock \emph{\bibinfo{journal}{\apj}} \textbf{\bibinfo{volume}{345}}, \bibinfo{pages}{245} (\bibinfo{year}{1989}).

\bibitem{Baldwin1981}
\bibinfo{author}{{Baldwin}, J.~A.}, \bibinfo{author}{{Phillips}, M.~M.} \& \bibinfo{author}{{Terlevich}, R.}
\newblock \bibinfo{title}{{Classification parameters for the emission-line spectra of extragalactic objects.}}
\newblock \emph{\bibinfo{journal}{\pasp}} \textbf{\bibinfo{volume}{93}}, \bibinfo{pages}{5--19} (\bibinfo{year}{1981}).

\bibitem{Kewley2001}
\bibinfo{author}{{Kewley}, L.~J.}, \bibinfo{author}{{Dopita}, M.~A.}, \bibinfo{author}{{Sutherland}, R.~S.}, \bibinfo{author}{{Heisler}, C.~A.} \& \bibinfo{author}{{Trevena}, J.}
\newblock \bibinfo{title}{{Theoretical Modeling of Starburst Galaxies}}.
\newblock \emph{\bibinfo{journal}{\apj}} \textbf{\bibinfo{volume}{556}}, \bibinfo{pages}{121--140} (\bibinfo{year}{2001}).

\bibitem{Kauffmann2003}
\bibinfo{author}{{Kauffmann}, G.} \emph{et~al.}
\newblock \bibinfo{title}{{The host galaxies of active galactic nuclei}}.
\newblock \emph{\bibinfo{journal}{\mnras}} \textbf{\bibinfo{volume}{346}}, \bibinfo{pages}{1055--1077} (\bibinfo{year}{2003}).

\bibitem{Easeman2024}
\bibinfo{author}{{Easeman}, B.}, \bibinfo{author}{{Schady}, P.}, \bibinfo{author}{{Wuyts}, S.} \& \bibinfo{author}{{Yates}, R.~M.}
\newblock \bibinfo{title}{{Optimal metallicity diagnostics for MUSE observations of low-z galaxies}}.
\newblock \emph{\bibinfo{journal}{\mnras}} \textbf{\bibinfo{volume}{527}}, \bibinfo{pages}{5484--5502} (\bibinfo{year}{2024}).

\bibitem{Pilyugin2016}
\bibinfo{author}{{Pilyugin}, L.~S.} \& \bibinfo{author}{{Grebel}, E.~K.}
\newblock \bibinfo{title}{{New calibrations for abundance determinations in H II regions}}.
\newblock \emph{\bibinfo{journal}{\mnras}} \textbf{\bibinfo{volume}{457}}, \bibinfo{pages}{3678--3692} (\bibinfo{year}{2016}).

\bibitem{Dopita2013}
\bibinfo{author}{{Dopita}, M.~A.}, \bibinfo{author}{{Sutherland}, R.~S.}, \bibinfo{author}{{Nicholls}, D.~C.}, \bibinfo{author}{{Kewley}, L.~J.} \& \bibinfo{author}{{Vogt}, F. P.~A.}
\newblock \bibinfo{title}{{New Strong-line Abundance Diagnostics for H II Regions: Effects of {\ensuremath{\kappa}}-distributed Electron Energies and New Atomic Data}}.
\newblock \emph{\bibinfo{journal}{\apjs}} \textbf{\bibinfo{volume}{208}}, \bibinfo{pages}{10} (\bibinfo{year}{2013}).

\bibitem{Marino2013}
\bibinfo{author}{{Marino}, R.~A.} \emph{et~al.}
\newblock \bibinfo{title}{{The O3N2 and N2 abundance indicators revisited: improved calibrations based on CALIFA and T$_{e}$-based literature data}}.
\newblock \emph{\bibinfo{journal}{\aap}} \textbf{\bibinfo{volume}{559}}, \bibinfo{pages}{A114} (\bibinfo{year}{2013}).

\bibitem{Curti2017}
\bibinfo{author}{{Curti}, M.} \emph{et~al.}
\newblock \bibinfo{title}{{New fully empirical calibrations of strong-line metallicity indicators in star-forming galaxies}}.
\newblock \emph{\bibinfo{journal}{\mnras}} \textbf{\bibinfo{volume}{465}}, \bibinfo{pages}{1384--1400} (\bibinfo{year}{2017}).

\end{thebibliography}

\bmhead{Acknowledgements}
We thank the DESI and IllustrisTNG collaborations for providing the data used in this study. 

\bmhead{Funding}
E.W., C.L., Z.C., C.M., Y.C., H.Y., and X.K. discloses support for the research and publication of this work from the National Science Foundation of China [12473008], the Start-up Fund of the University of Science and Technology of China [KY2030000200], and the Cyrus Chun Ying Tang Foundations. C.L. discloses support from the Fundamental Research Funds for the Central Universities [WK2030250123] and the China Postdoctoral Science Foundation [2026T190848].

\bmhead{Author contributions} C.L. developed the initial concept, led the research effort, performed data reduction and scientific analysis, and wrote the manuscript. E.W. conceived the initial concept, initiated the project, and supervised the analysis. Z.C. contributed to the measurement of galaxy geometry. C.M. contributed to the data acquisition and processing of data in cosmological simulations. Y.C., H.Y., and X.K. contributed to the interpretation of the results and the conceptual framework. All authors discussed the findings and commented on the final manuscript.

\bmhead{Materials \& Correspondence} Correspondence and requests for materials should be addressed to E.W. (ecwang16@ustc.edu.cn).

\bmhead{Competing Interests} The authors declare no competing interests.

\clearpage

\begin{appendices}
\end{appendices}

\end{document}